\documentclass[fleqn,10pt]{wlscirep}
\usepackage{enumitem}
\usepackage{graphicx}
\usepackage{multirow}
\usepackage{booktabs}
\usepackage{rotating}
\usepackage{amsmath}
\usepackage{amsfonts}
\usepackage{todonotes}
\usepackage{subcaption}

\usepackage[utf8]{inputenc}

\title{Undermining and Strengthening Social Networks through Network Modification}
\author[1,2]{Jonathan Mellon}
\author[1,3,*] {Jordan Yoder}
\author[1]{Daniel Evans}
\affil[1]{Network Science Center, United States Military Academy, West Point, NY }
\affil[2]{Nuffield College, University of Oxford, Oxford, United Kingdom}
\affil[3]{Applied Mathematics \& Statistics, The Johns Hopkins University, Baltimore, MD}
\affil[*]{Corresponding author: jyoder6@jhu.edu}

\begin{document}

\begin{abstract}
Social networks have well documented effects at the individual and aggregate level. Consequently it is often useful to understand how an attempt to influence a network will change its structure and consequently achieve other goals. We develop a framework for network modification that allows for arbitrary objective functions, types of modification (e.g. edge weight addition, edge weight removal, node removal, and covariate value change), and recovery mechanisms (i.e. how a network responds to interventions). The framework outlined in this paper helps both to situate the existing work on network interventions but also opens up many new possibilities for intervening in networks. In particular use two case studies to highlight the potential impact of empirically calibrating the objective function and network recovery mechanisms as well as showing how interventions beyond node removal can be optimised. First, we simulate an optimal removal of nodes from the Noordin terrorist network in order to reduce the expected number of attacks (based on empirically predicting the terrorist collaboration network from multiple types of network ties). Second, we simulate optimally strengthening ties within entrepreneurial ecosystems in six developing countries. In both cases we estimate ERGM models to simulate how a network will endogenously evolve after intervention.
\end{abstract}

\flushbottom
\maketitle

\thispagestyle{empty}

\section{Introduction}

An ever increasing number of studies have documented how individuals' behavior is influenced by their social context. Network structure has been linked to individual and collective outcomes in contexts including criminal networks \cite{morselli2013crime, scott2011sage, pyrooz2012continuity}, workplaces
\cite{tsang2012nursing, jokisaari2013role, zou2013bonds, venkataramani2013positive}, classrooms
\cite{russo2005prestige, huitsing2012must, lomi2011some} and industries
\cite{mcevily2012not, lavie2007alliance, koka2008designing}. Given the effect that network structures have on outcomes, it is often useful to understand how attempting to externally influence a network is likely to change its structure and whether an attempted intervention will achieve its goals.

Most of the work on intervening in networks has focused on two versions of the key player problem: 1) which nodes to remove in order to minimize network cohesion (KPP-NEG) and 2) which nodes to use as seeds to maximize diffusion in a network (KPP-POS) \cite{borgatti2006keyplayers}. While these problems have important applications, we argue that they are part of a more general class of network intervention problems that can be categorized according to their outcome metric, intervention strategy and the response of the network to the intervention. 

We outline a general framework for network modification that allows for arbitrary objective functions, types of modification (e.g. edge weight addition, edge weight removal, node removal, and covariate value change), and recovery mechanisms (i.e. how the network responds to interventions). We show how existing work fits into this framework and how variations within the framework can lead to new applications for these techniques and alternate solutions to existing applications. 

Many problems in this wider framework have not been deeply studied. In particular, strategies that involve modifying ties rather than nodes of a network have received relatively little attention. This is despite the fact that many potential interventions in a network have effects at the tie level (e.g. assigning people to work together or signing agreements) rather than the node level. Similarly, most studies have not focused on how a network will respond to an intervention. This is a potentially important oversight given that a recent study suggested that network response to interventions would render most interventions ineffective \cite{duijn2014relative}. Finally, there has been little work on defining outcome metrics empirically in terms of how they affect ultimate objectives such as terrorist attacks, level of drug supply, or student outcomes. 

In this paper, we seek to organize the various strategies for network modification in a singular framework.  Then, we demonstrate the use of this framework in two case studies on real world data to model how objective function and recovery mechanisms can be empirically derived from the problem--rather than being posited on theoretical grounds--and show an applied example of optimizing edge weight additions. 

First, we look at the case of the Noordin terrorist group. Rather than defining the metric function a-priori, we empirically calibrate it by using the observed network to predict a terrorist collaboration network using a multiple regression quadratic assignment procedure model (MRQAP). The predictions from this model then form the basis of the metric function that we try to reduce. We also show how the network is likely to evolve in response to such an intervention using an exponential random graph model fitted to the initial network as a network recovery algorithm. Second, we look at networks of roles in the entrepreneurial ecosystems across six cities in developing countries. We use this case study to demonstrate how the framework can be applied to interventions involving strengthening ties, where the aim is to make the network perform more effectively. In this case, we look at what ties we would try to strengthen in the poorly performing ecosystems to make them more similar to the best performing ecosystem (of the six studied): Accra in Ghana. Finally, we conclude with a discussion of further possible applications and modifications to the framework in the paper.

\section{Methods} The network modification framework has four components: an outcome metric, a strategy, an optimization approach, and a network evolution mechanism. The outcome metric, strategy, and evolution mechanism constitute the problem, while the optimization approach is the solution.

 The methodology can be summarized as:\begin{enumerate}
\item Begin with an initial network (weighted or binary adjacency matrix) and a metric function to be minimized as determined by the goal of the modification
\item Propose changes of a selected type (e.g. node removal) to the initial network.
\item Optimize choices of changes (within specified constraints such as a budget) according to an optimization method
\item Assess the extent to which network interventions will persist by simulating endogenous evolution of the network
 \end{enumerate}

\subsection{Choosing a metric function}

Generally, one who desires a network to change has a preferred outcome.  Graph metric functions (which are not necessarily mathematical metric distance functions) are objectives that seek to capture this preferred  outcome in quantitative terms. In this paper, we call a network a pair of sets $G = (V, E),$ where $G$ is the network, $V$ is a set of nodes, and $E$ is a set of edges.  We define a graph metric function to a be function $$f : \mathcal{G} \rightarrow \mathbb{R},$$ where $\mathcal{G}$ is the set of all networks.  Lower values of the function $f$ are more desirable.Choosing a metric function is a task that should be specific to the actual application.

Some previous work on network modification has focused on network features (such as centrality or brokerage statistics) themselves with the implicit assumption that these will further an ultimate goal \cite{roberts2011strategies}. In other work, a link to an ultimate goal is made more explicit. For instance, the KPP-NEG approach aims to leave the network with the least possible cohesion \cite{borgatti2006keyplayers}. Another explicit optimization criteria was set forth in \cite{ortiz2008information}, where they use information theory to formulate an objective function.  Also, Duijin et al. \cite{duijn2014relative} use a custom efficiency measure. However, in this case, the measure is problematic as it typically attains worse values when members of a network are removed, even in the absence of a network recovery mechanism. In fact, the measure will be worst when there are no members of the network remaining. It is therefore implausible that this statistic is a useful measure of a criminal network's effectiveness. This strange behavior (removing nodes improves the efficiency of the network) will likely not hold in practice.  Due to this discrepancy, improving the chosen graph metric could end up giving very suboptimal results.

One approach that has not been previously taken is to empirically tie the objective function to an ultimate objective. For example, "the  predicted cocaine production of the network'' is a quantitative measure of the efficiency of the network that we might wish to disrupt.  We suggest calibrating the metric against that outcome empirically. For instance, we could look at what network structures predict an outcome such as total drugs produced by a criminal network, overall student attainment in a high school, or total sales in a department. By empirically predicting an ultimate outcome, we can simultaneously combine multiple network statistics or even networks themselves into a quantitative measure demonstrably associated with the desired outcome.

\subsection{Choosing Allowable Types of Changes}
There are many different types of interventions that it is possible to make to a network. However, these can be summarized into five categories: node removal, node addition, tie weight addition, tie weight removal, and node covariate change. 

The most common type of intervention in the literature has been node removal, particularly in terms of disrupting either the network itself \cite{borgatti2006keyplayers, duijn2014relative, ortiz2008information} or contagion through a network \cite{arulselvan2009detecting}). After this, the next most common form of intervention is node covariate change. The KPP-POS problem can be formulated in terms of optimizing covariate change. From this perspective, we optimize a metric of ``reachability from tagged nodes'' through the strategy of choosing which nodes to tag (i.e. changing a node's covariate value of being tagged).  However, covariate change could be optimized in many other ways. For instance, if a network has a tendency to form homophilous ties, changing a covariate value could be an effective way of reorganizing the network structure itself. 

After node removal, edge modification is the next natural choice.  However, it has been studied much less thoroughly.  Edge deletion has been a focus of previous research in this area.  \cite{girvan2002community} set forth a method in which edges are deleted, then a network property is recalculated in order to assess a particular edge's importance to the network.  \cite{valente2010bridging} used a similar idea more recently to study bridging measures.

There has been little work so far considering the problem of how to optimize \emph{strengthening} edges or adding nodes into a network. Table \ref{tab:network_intervene} shows how various forms of network interventions could work in applied contexts. Although moving a child to a different classroom, firing an employee, and arresting a criminal appear to be very different substantively, they are fundamentally the same from a network perspective. The potential applications are almost limitless but could include making strategic hires in academic departments based on citation networks, decreasing segregation in classrooms and making alliance decisions to maximize trade.

As well as defining what types of changes to the network will be considered, we also need to define the constraints on what changes can be made. The simplest case of this is a budget that limits the number of changes that can be made. However, a more complex set of constraints is also possible (e.g. add a maximum of 50 total edge weight, with no more than 10 going to a single edge). In this case it would also be possible to assign different costs to different changes (e.g. removing node $v_1$ from the network costs 5 but node $v_2$ only costs 2 to remove). The examples in this paper only consider the case of a simple budget of $X$ changes of a single type. This budget can be viewed in terms of time, money or simply the social limits of how much change can be achieved by an intervener.

\subsection{Optimization Method}
Consider the problem $\min_{G \in \mathcal{G} f(G)}$ subject to zero or more constraints.  For most graph metric functions, this discrete problem will be quite challenging.  General purpose discrete optimization strategies generally revolve around approximation guarantees or heuristics.

We use two optimization strategies in this paper. The simplest strategy is exhaustive search, where all possible combinations of node removals are considered and rated. This method is guaranteed to reach the optimal solution for a discrete optimization problem because every solution is considered.  For the same reason, it is completely impossible to implement on graphs of large size.

The other methodology we focus on in this paper is greedy optimization. Greedy algorithms are designed around the greedy heuristic, where the locally optimal choice is made at each step.  Sometimes, this will lead to poor results.  Other times, it works well enough.  In both cases, using a greedy heuristic can be a massive speedup over brute force techniques such as exhaustive search or dynamic programming.  Due to the inherent hardness of discrete optimization on graphs in general, it is no surprise that the greedy heuristic has been used recently for graph modification purposes \cite{borgatti2006keyplayers, valente2010bridging, ortiz2008information}.

\begin{figure}[ht!]
\centering
\includegraphics[width=10cm]{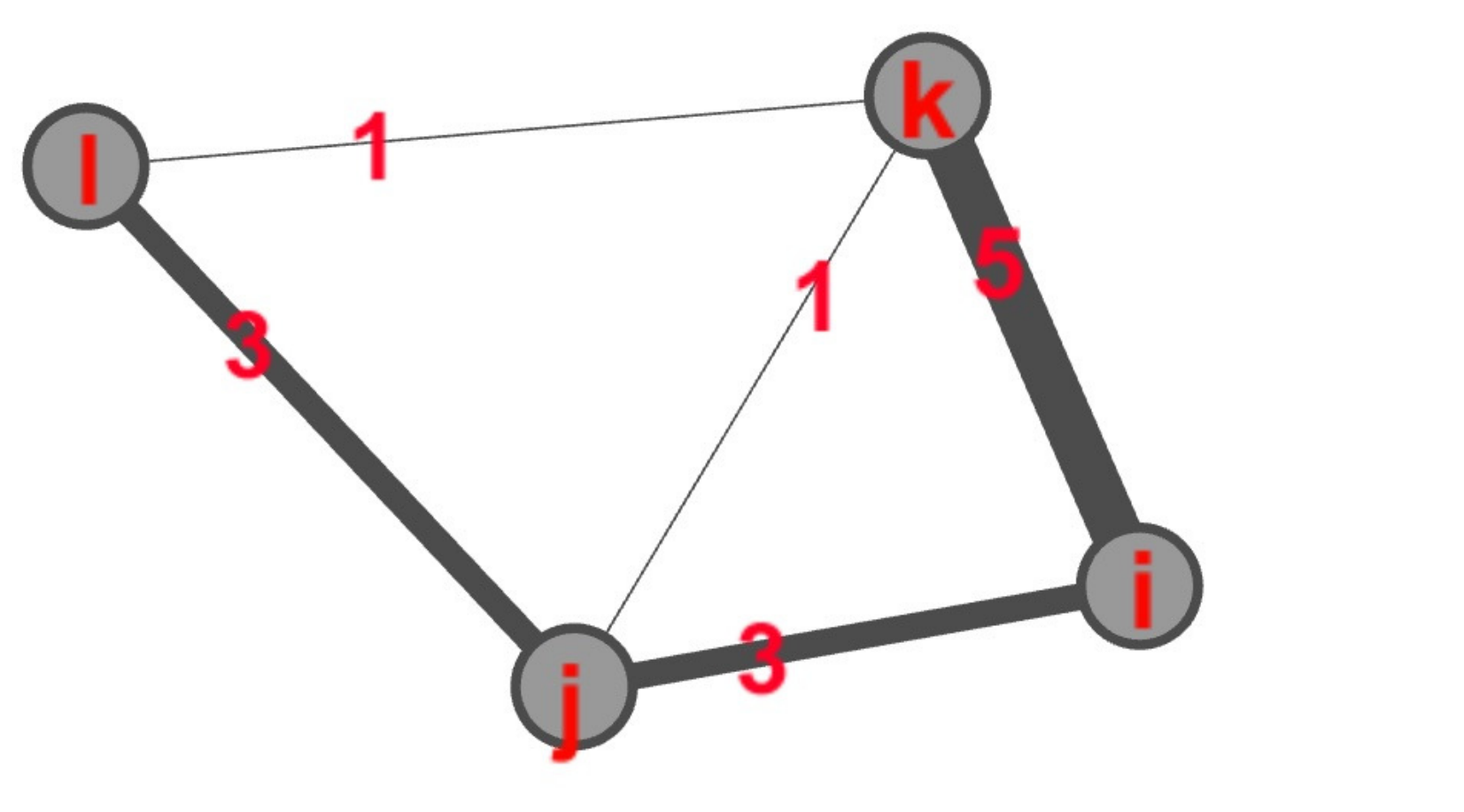}
\caption{An example of a network where greedy optimization will not find the optimal solution. Note that all ties can be removed if \textit{j} and \textit{k} are removed but that \textit{i} has the most edge weight in the first round.}
\label{fig:counterexample}
\end{figure}

There are classes of problems (greediods) for which a greedy algorithm is provably optimal \cite{oxley2006matroid, helman1993exact}, however even the relatively simple outcome measure in this paper is not one of these problems. Figure \ref{fig:counterexample}  shows an example of a network where the greedy algorithm will choose to remove nodes \textit{i} and \textit{j} rather than the optimal set of \textit{j} and \textit{k}. Such counterexamples can be generated using linear programming to find edge weight combinations that will mislead a greedy algorithm. Nevertheless, greedy optimization will find a good or even optimal solution in many real network modification examples.

\subsection{Simulating endogenous network change}
The final step of the framework is to consider how the network will respond to an intervention. Most studies omit this step entirely \cite{borgatti2006keyplayers, everton2012Topography}, implicitly assuming that the network will not change in response to the change. However, a few studies have included this extra step. Recently, Duijin et al.\cite{duijn2014relative} conducted a detailed recovery study by specifying several possible mechanisms through which a network might recover. Unfortunately, their recovery mechanisms are somewhat unrealistic as they essentially involve adding random ties into the network and ignore the existing structure and what mechanisms were likely present that generated it originally.  In \cite{liu2012criminal}, they use game theoretic considerations for recovery mechanisms. 

Rather than specifying a network recovery algorithm \emph{a priori}, we empirically calculate a mechanism by using an exponential random graph model (ERGM) to calibrate an underlying generative process within the network. This means that we assume that there is an underlying process that creates the structure within a network and that this mechanism will continue after an intervention takes place. This assumes that an intervention is able to affect the strength of ties in a network but does not directly change the underlying social processes within the network.

Because we use weighted networks, it may be useful to revisit the notations of ERGMs to account for this.  For a more rigorous construction, see \cite{krivitsky2012exponential}.  Consider the adjacency matrix of a network as a random variable $Y,$ where $Y_{i,j}$ is a non-negative random variable representing the weight of the tie between vertices $i$ and $j$.  Ties are hypothesized to depend on one or more statistics of a given configuration of the networks: $s(A),$ where $A$ is a realization of $Y$ and $s$ is possibly vector valued.  There is a parameter, $\theta$, which is also possibly vector valued that is estimated as part of the fitting process.  Finally, we specify a reference measure $h(A)$ which can loosely be interpreted to give the prior probability of a given configuration $A$.  Putting it all together,
$$P(Y=y) = \frac{h(y)\exp\left(\theta^T s(y)\right)}{\sum_A h(A) \exp\left(\theta^T s(A)\right)}.$$

\section{Case Study 1: Noordin}
To demonstrate empirically calibrating an objective function we apply our framework to a dataset of relationships in the Indonesian Noordin terrorist group.  The Noordin dataset was constructed \cite{roberts2011strategies} based on a 2006 International Crisis Group report on Noordin Mohammed Top's terrorist network in Indonesia \cite{ICGNoordinReport2006}. Noordin was involved in the 2002 Bali bombing and the 2003 bombing of the Marriot hotel in Jakarta, as well as several later attacks. The dataset contains extensive information about 79 terrorists and their connections to each other both in terms of one-mode networks (communication and friendship) and two-mode networks (joint membership of a mosque or terrorist organization). Importantly, the dataset also includes information on which terrorists were directly involved in each attack. This allows us to create a collaboration network of terrorists that shows whether they are jointly involved in an attack. 

The Noordin dataset has previously been used to examine network disruption. However, previous analyses have taken a heuristic approach to disruption by comparing the implications of adopting kinetic and non-kinetic disruption strategies \cite{roberts2011strategies} and examining the network's topography \cite{everton2012Topography}.

\subsection{Noordin: Graph Metric}

Disruption of a terrorist network is conducted primarily for the purpose of reducing the effectiveness of a terrorist network. We therefore choose an outcome measure that is directly related to this. We define a terrorist network's effectiveness in terms of the number of attacks that terrorists collaborate on. Six terrorist attacks in the Noordin dataset involve multiple terrorists. We can therefore create a terrorist attack collaboration network that shows which terrorists were co-involved in different attacks. We normalize this network so that each attack contributes a total of 1 edge weight. Since, we would like to disrupt a network prior to attacks taking place, we aim to reduce the number of expected attacks from a network. We therefore use the other ways in which terrorist are related: communication ties, organizational ties and educational ties to predict which terrorists are most likely to collaborate with each other. 

We therefore aim to optimize the total predicted edge weight of the terrorist collaboration network based on a MRQAP model using communication, organizational and educational ties. Because of the normalization of edges, this metric corresponds to the expected number of attacks by the network. The MRQAP metric also has an another appealing property: because all the coefficients are positive, removing a terrorist will never increase the expected number of terrorist collaborations (in the absence of network recovery).

We use a quasi-likelihood method with a log-link using the {\bf glm()} function in R and the quasipoisson family.  That is, we model $$\log(\mu) = \beta_0 + \beta_1 x_1 + \dots + \beta_q x_q,$$ for some scalars to be fit $\beta_i$ and independent variables $x_i.$   Further, if $Y$ is our response variable, we assume $\mathbb{E}[Y] = \mu.$  The ``quasipoisson'' family assumes that $Var(Y) = \phi \mu$ for some scaling parameter $\phi,$ which is not necessarily unity.  Note that nothing about these assumptions implies that the response variable need be discrete. The parameters are fitted using quasi-maximum likelihood (QMLE).

The usual calculations for standard errors are biased because of autocorrelation within rows and columns of an adjacency matrix. The MRQAP approach calculates p-values through a permutation test that uses the dataset to assess how often the estimated t-statistic would have been greater than the observed t-statistic under a simulated null hypothesis. In order to account for the dependency within rows and columns, the permutations simultaneously permute the labels for rows and columns. This maintains the same dependency structure within individuals in the dataset but still gives us an estimate of the natural variance in the data. We use the double semi-partialling approach \cite{dekker2007sensitivity}. This has been shown to be the permutation approach that most reliably approximates the correct type I error rate, while retaining good power. The results from a MRQAP model are read like a GLM table, with the parameters showing how a one unit change in the independent variable changes the log of the expectation of the dependent variable, and the p-value showing us how often we would expect to see a parameter of this magnitude under the null hypothesis.

We find that three factors are strong predictors of terrorist collaboration in the Noordin network: communication, educational ties and organizational ties. While the variance explained in the MRQAP model is moderate (mostly due to the sparsity of the collaboration network), but it still represents a substantial improvement over previous network metrics, which have not been demonstrated to be correlated with important outcomes at all. All three networks are positively related to terrorist collaborations, but only the communication and education networks reach significance as we can see in table \ref{tab:mrqapoutcome}.

With the regressions complete, we formulate the objective function.  For a given node set that has been selected for removal, say $R$, our metric function is (up to a constant),
$$f(R) = \sum_{i, j : i < j} \mathbb{E}[Y_{i,j}]\chi_{\{i,j \not\in R\}},$$
where $$\chi_{A}:= \begin{cases} 1 \mbox{ if } A \\ 0 \mbox{ else} \end{cases}$$ is the indicator function for event A.

\subsection{Noordin: Allowable Changes}
As with most previous analyses of dark networks, we focus on node removal. While other modifications may be possible (sowing dissent between factions or introducing undercover agents for instance), node removal is the most realistic intervention that outsiders can make to a terrorist network by arresting members of the group. While our model assumes that the intervener has some ability to affect the network structure, we assume that this ability is not limitless. As a result, we reflect their options through an intervention budget of changes they can make. For this paper, we assume that one budget unit corresponds to the ability to remove one node from the network. For the experiments in this paper we use an intervention budget of 5 and 15. This constitutes a modest intervention in the network, which contains a total of 79 nodes.

\subsection{Noordin: Optimization strategies}
As previously mentioned, we focus on greedy optimization, where 1 node is considered at a time for removal.  The node reducing the graph metric the most is removed.  Then, all remaining nodes are considered again for removal.  This is considerably faster than the exhaustive search, which looks at all subsets of nodes of size no more than the budget.  For a budget of 5 removals, this is plausible to run, but for 15 removals, we estimate that the algorithm would take around 3.2 years to run.  Further work on the algorithm and distributing the computation would substantially reduce this time, but the general observation that exhaustive search becomes rapidly intractable with larger numbers of decisions undoubtedly remains.

\subsection{Noordin: Endogeneous Network Change mechanism}
We fit a simple ERGM using the covariates available.  We model the communication network as a function of the number of edges, number of isolates, geometrically weighted edgewise shared partner distribution, and as dependent on other networks.  That is, we consider the organizational and educational networks as exogenous within the timeframe we are examining.  Future work could look at co-evolution models which would allow both communication and organizational networks to evolve together. However, this is beyond the scope of this paper.

To model the dependency of the Noordin communication network on other networks between these individuals, we include the sum of each of the following covariate values in the network in the ERGM: organizational co-affiliation, shared educational background, friendship, kin and religious co-affiliation (for instance sharing a mosque). A separate statistic is included in the ERGM equation for each of these dyadic covariate networks. The statistic for each is the sum of the covariate values across dyads that have a communication tie present.  In addition to these networks, we also model the communication endogenously with several structural effects. The simplest of these is simply the network's ties, which we sum across the whole network in the edges statistic. Additionally, we include the number of isolates as a statistic in the ERGM equation. We also model triadic closure by including  the geometrically weighted edgewise shared partner (GWESP) distribution, defined as:

\begin{equation}
w = e^\alpha \sum_{i=1}^{n-2}\{ 1 - (1 - e^{-\alpha})^i \}p_i, 
\end{equation}

where $p_i$ is the number of pairs of actors who are connected with a tie and are tied to exactly $i$ of the same actors, $n$ is the total number of actors in the network and $\alpha$ is a constant that controls how much more weight pairs of actors with many acquaintances in common are given in the statistic. For instance, when $\alpha=0.5$ a pair of actors with one acquaintance in common receive a weight of 0.61 and a pair of actors with two acquaintances in common receive a weight of 0.85 and a pair with 5 joint acquaintances receives 0.99. When $\alpha=0.25$, these weights are 0.78, 0.95, and 1. When $\alpha=0$, these pairs are all given the same weight of 1. Following \cite{goodreau2009homophily}, we set $\alpha=0.25$.

The ERGM results in table \ref{tab:ergm} show that the terrorist communication network can be modeled as arising from the organizational co-affiliation, friendship, kin and religion networks. Interestingly, the shared education network doesn't predict communication ties after accounting for these other networks. As expected, the GWESP statistics show that the communication network has a strong tendency towards closing triads.

\subsection{Noordin: Experiment results}
  To summarize the previous sections, we estimate the total number of attacks as the sum of the upper triangular of matrix $\mathbb{E}[Y]$ (since all covariate networks are and will be symmetric, we can cut everything but the upper triangle). We will allow removal of nodes.  Here, we maintain the size of the networks by zeroing all covariates to indicate that an inexperienced, unconnected replacement has been found.  This is slightly generous towards the terrorists, as it implies that their recruitment team is very responsive to forced downsizing.  These inexperienced terrorists contribute a mere $1.33e-3$ attacks in expectation each.

We run our greedy algorithm with a budget of 15 removals considered one at a time.  For comparison, we performed an exhaustive search for 5 removals. We compare the results from the greedy algorithm to the natural heuristic of removing the most influential actors (as determined by degree), which has been used in previous studies of network disruption \cite{duijn2014relative,roberts2011strategies}.

\begin{figure}[ht!]
\centering
\includegraphics[width=10cm]{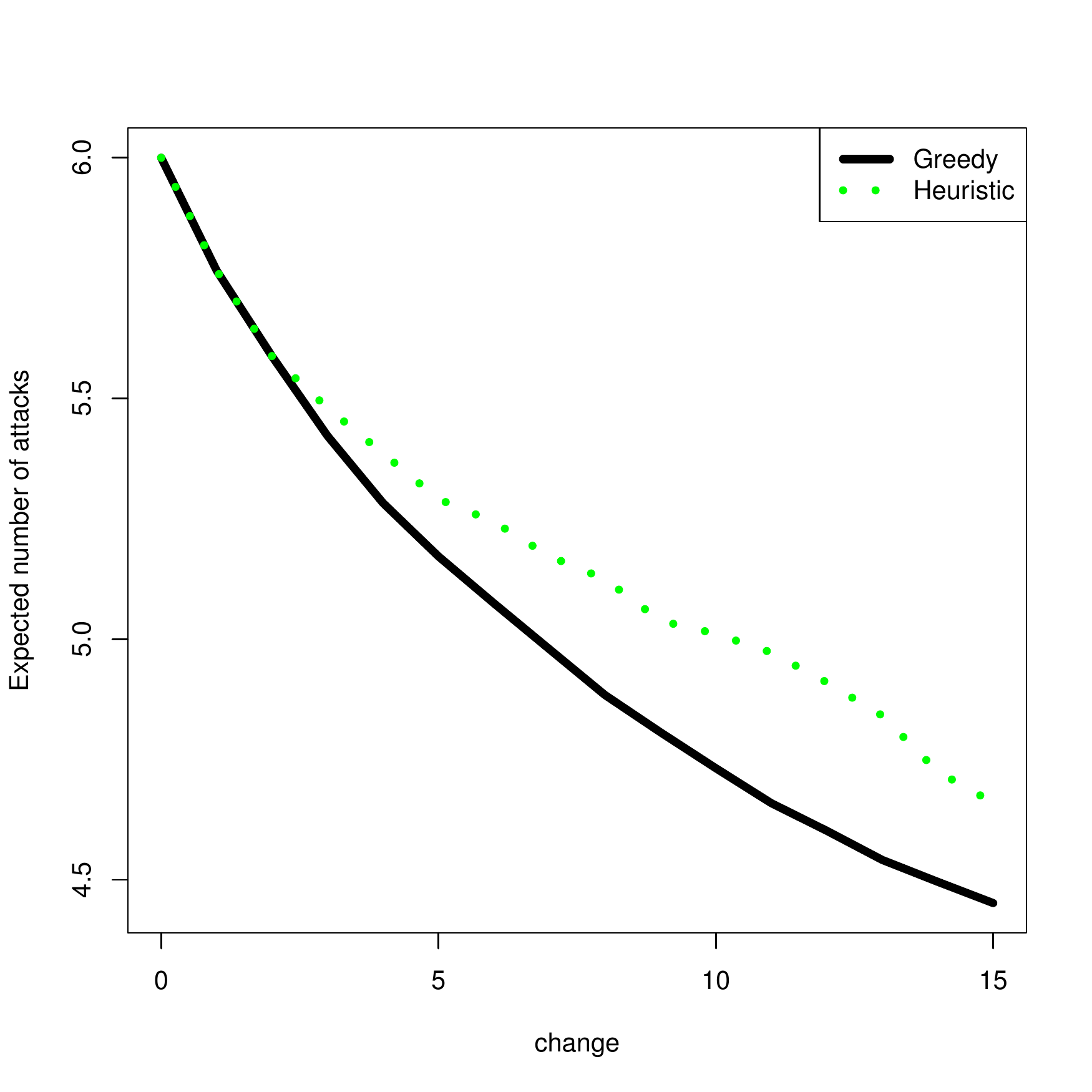}
\caption{The number of expected attacks as a function of total removals allowed for the greedy and a heuristic algorithm. }
\label{fig:experimentOptimality}
\end{figure}

We use the fitted ERGM to simulate how the metric will degrade (recover) over time after performing removal using the greedy and heuristic strategies at either 5 or 15 removals   (cf. Figure \ref{fig:experimentOptimality}).  The heuristic algorithm improves the metric, but lags behind the greedy algorithm after the first few removals.  Because the run-time of the greedy algorithm was less than two seconds, we do not consider this a large consideration when choosing between algorithms. 

\begin{figure}[ht!]
\centering
\includegraphics[width=10cm]{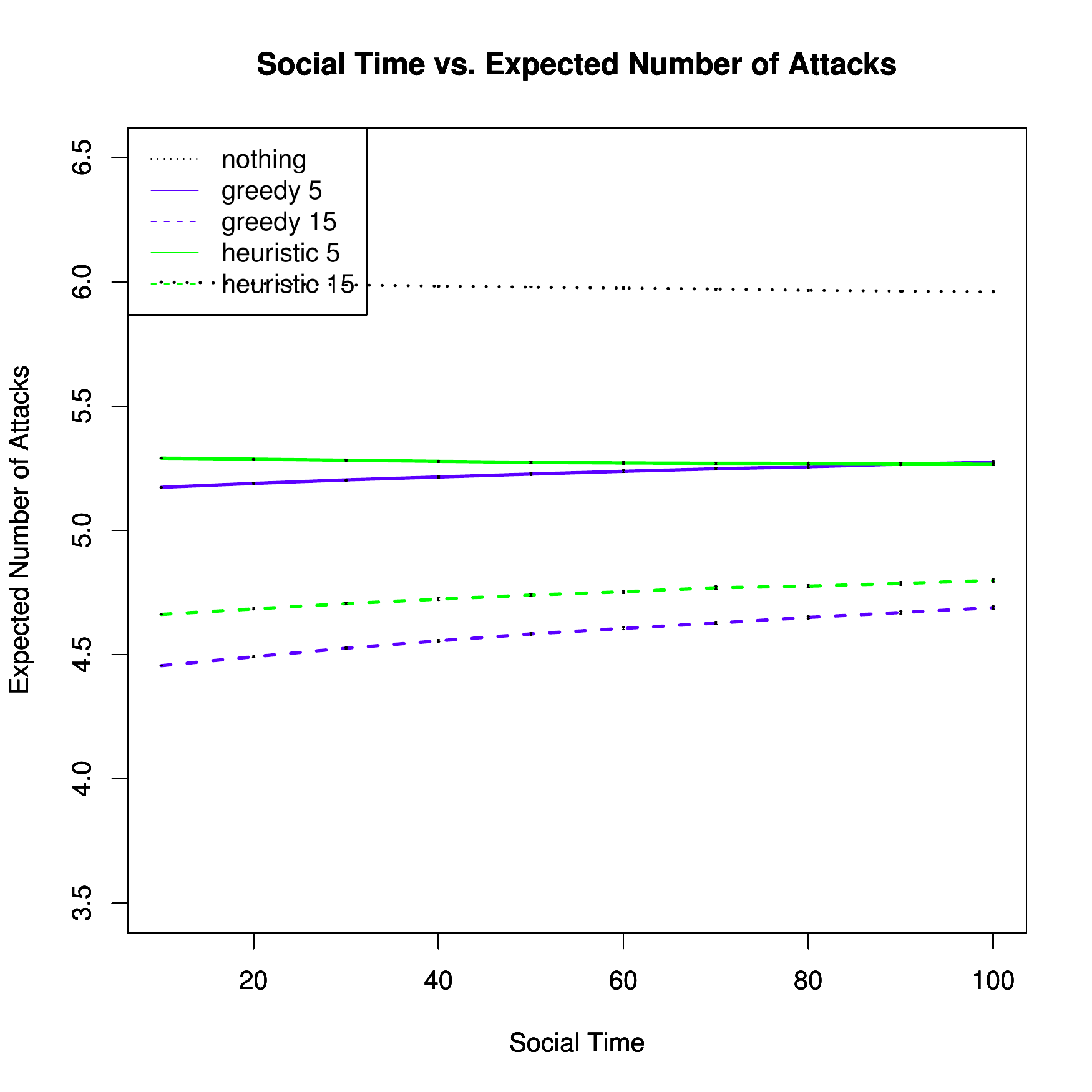}
\caption{The simulated number of expected attacks as a function of social time.  Lower is better.  Observe that the greedy strategy consistently outperforms the heuristic strategy.}
\label{fig:dissEvolution}
\end{figure}

Using the trained ERGM, we simulate the evolution of the modified networks resulting from the removals of nodes as suggested by each strategy. We observe that the greedy method performs the best both initially and across the full 100 steps of social time that the ERGM runs (on average across 1000 sampled paths) (cf. Figure \ref{fig:dissEvolution}). This shows that greedy optimization both outperforms the heuristic approach and maintains that performance after the network endogenously recovers.

\section{Case Study 2: Strengthening Economic Networks in the EntrepEco Dataset}
In a recent technical report, \cite{daniel2014methodology} introduces the EntrepEco (Entrepreneurial Ecosystems) Dataset.  Six cities with burgeoning economic networks were studied (Addis Ababa, Dar es Salaam, Monrovia, Lusaka, Accra, and Kampala).  Entrepreneurs were surveyed through snowball sampling about which of 13 types of resources they would ask for help given a particular problem (self, government representative, government business development, incubator, NGO, investor, family, religious, social network, bank, professional, military, and education).  Then, a bipartite network for each city is created with links between entrepreneurs and the resources.  We use a modified version of the EntrepEco Dataset in which we excluded responses that involved ``self," as  self has a different meaning depending on the context, which would complicate the next step if not accounted for.   Thus, our bipartite graphs included 12 roles.  This bipartite network was then forced into a weighted unimodal network by connecting resources with a common entrepreneur.  

We identified Accra as the most vibrant network of all the cities studied.  This is based on our experience, and several articles corroborate this (cf. \cite{vc4a2015,  nam2015}). Thus, we now attempt to modify the networks of the other cities to be more like Accra's.

\subsection{EntrepEco: Graph Metric}
Because Accra was identified as the strongest entrepreneurial network, we want to maximize the similarity to that network.  We can capture that desire in a graph metric function:
if $A^{(j)}$ is the adjacency matrix for the $j^{th}$ city, we used the negative cosine similarity between $A^{(j)}$ and $A^{(Accra)}:$
$$f(A^{(j)}; A^{(Accra)}) = 1-\frac{\sum_{i,k}A^{(j)}_{i,k}*A^{(Accra)}_{i,k}}{\|A^{(j)}\|_F \|A^{(Accra)}_F},$$ where $\|A\|_F = \sqrt{\sum_{i,k}A_{i,k}^2}$ is the Frobenius norm of $A$.

\subsection{EntrepEco: Allowable Changes}
Unlike the Noordin case study, removing nodes is not a practical form of intervention when considering roles in an entrepreneurial ecosystem. However, it may be within the power of a decision maker to strengthen links between certain roles, for instance through subsidizing joint investments or facilitating meetings. For the experiments in this section we use an intervention budget of 50. This constitutes a substantial intervention in these cities (around a quarter of the existing total edge weight for each of these cases).

\subsection{EntrepEco: Optimization Method}
As with the first case study, we use greedy optimization to try to find the set of changes to edges that minimizes the distance between the starting network and the target network of Accra.

\subsection{EntrepEco: Endogenous Evolution Mechanism}
As with the first case study, we use ERGM models to simulate the network recovery process. We fit separate weighted ERGMs to each city, to model the underlying processes generating its network structure.  We include effects for non-zero edges, the sum of edge weights, total edges and the transitive weights which captures triadic closure. 

\subsection{EntrepEco: Experimental Results}

Table \ref{tab:allcitychanges} shows the distribution of edge weights that the algorithm gives to each edge for the other five entrepreneurial networks. In the two cities that were previously identified as most different from Accra, Addis Ababa and Monrovia, the algorithm spends its entire intervention budget on boosting the social network/professional/incubator role triad.  In the other cases, the algorithm tends to focus on ties involving incubators, investors, social networks and professionals (with the one exception of Kampala where a small amount of edge weight is assigned to increasing the strength of the government representative-social network tie).

\begin{figure*}[ht!]
\begin{subfigure}[b]{0.5\linewidth}
\centering
\includegraphics[width=.9\columnwidth, height = 0.9\columnwidth]{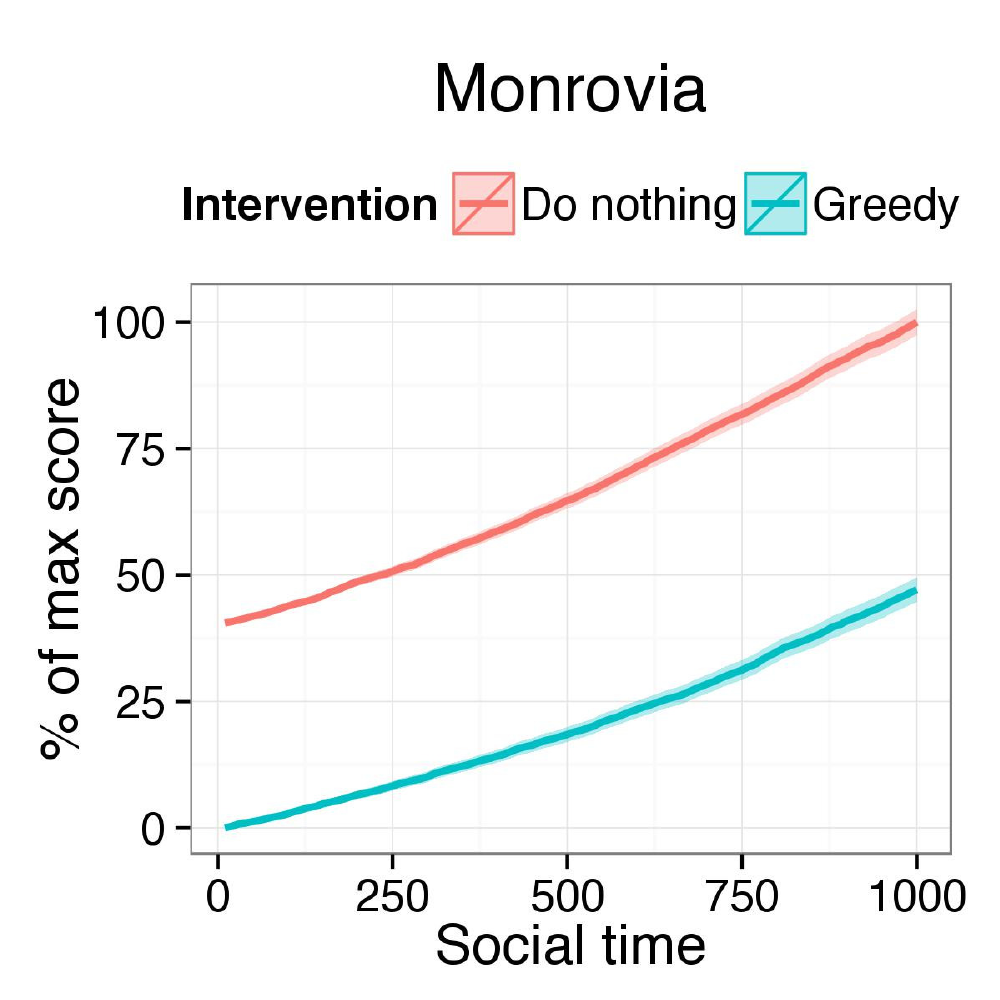}
 \label{fig:mo_intervene}
\end{subfigure}
\begin{subfigure}[b]{0.5\linewidth}
\includegraphics[width=.9\columnwidth, height = 0.9\columnwidth]{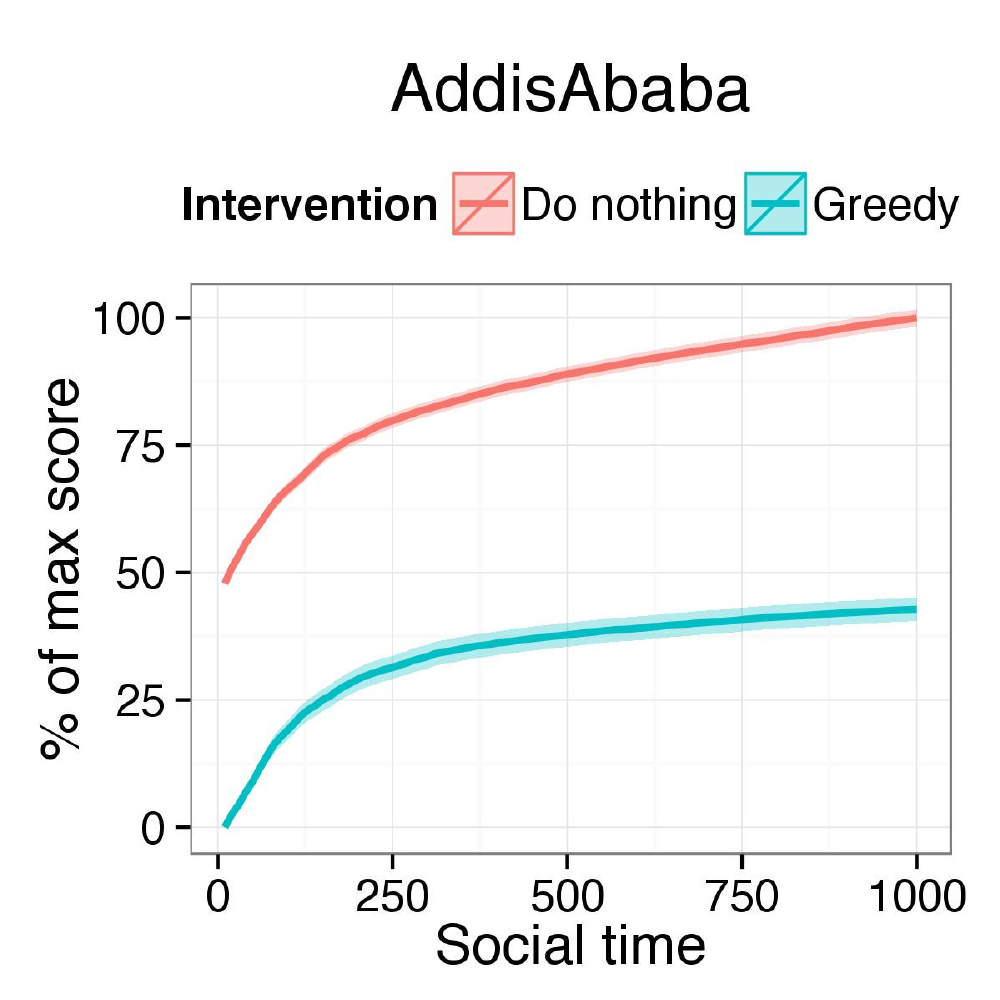}
 \label{fig:aa_intervene}
\end{subfigure}
\caption{Average optimization score for Monrovia (a) and Addis Ababa (b) across 1000 Monte Carlo Replicates for the evolution of the random metric of the proposed network originating from two methods: blue is do nothing and red is greedy.}
\label{fig:intervene_evolve}
\end{figure*}

In Figure \ref{fig:intervene_evolve}, we present a line plot of the average evolution of the values of the graph metric functions across 1000 Monte Carlo replicates using the fitted ERGM models to simulate the evolution of the Monrovia and Addis Ababa networks after intervention. Recall that lower values are preferred. Greedy selection greatly improves the network intervention effectiveness compared with doing nothing or picking the best random draw. The performance of all strategies degrades with time, as the endogenous network processes move the network away from the modified network.  For comparison, we included doing nothing as an alternative strategy.  We scaled the y-axis so they would always be on a 0-1 scale.  Lower is better.  Error shading is $\pm$2 standard deviations on the 0-1 scale, computing across Monte Carlo replicates. We did not include the other cities, as the plots look very similar.

\begin{figure}[h!]
   \centering      
 \includegraphics[width=\linewidth]{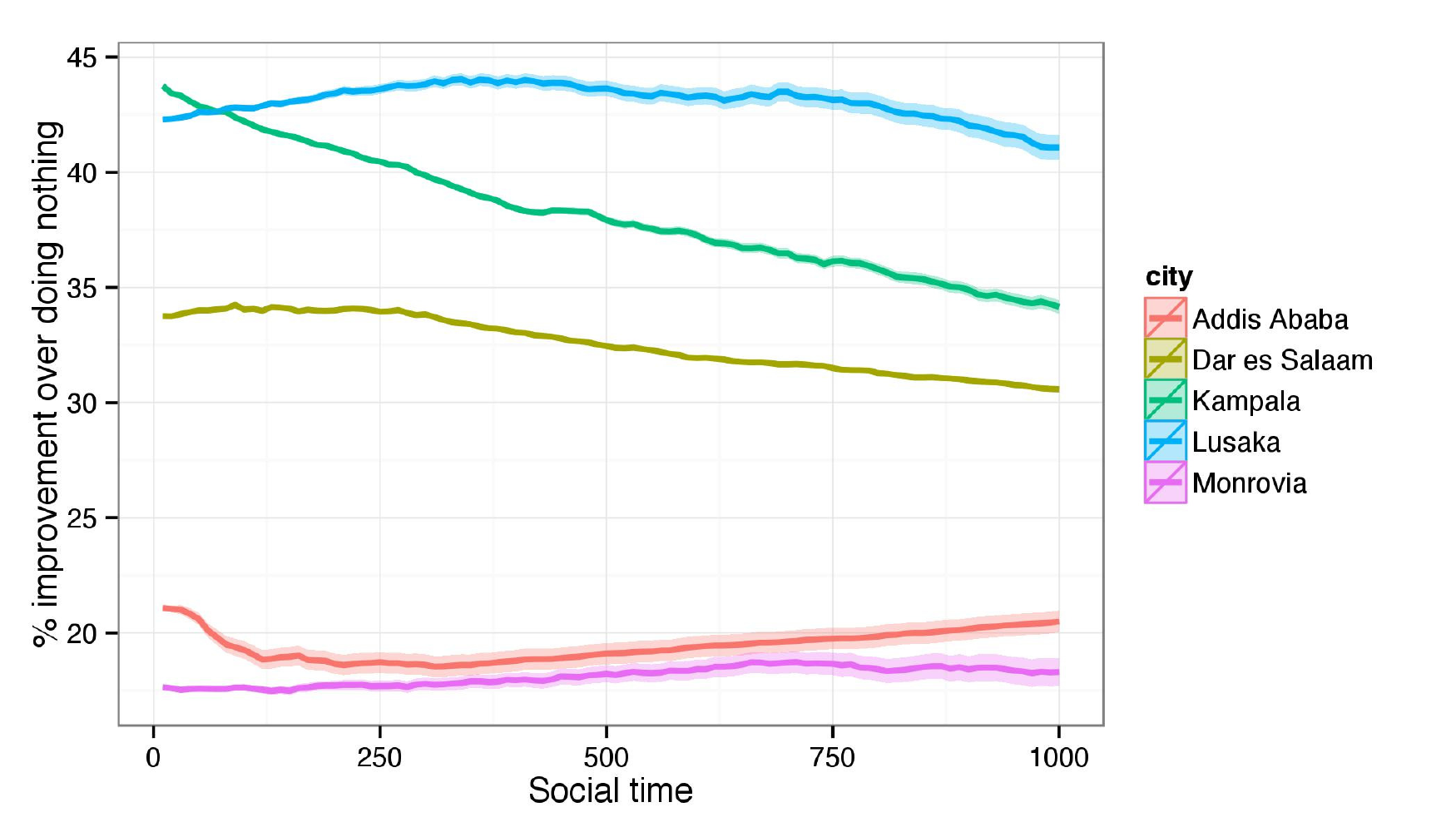}
 \caption{Average percentage improvement (in terms of similarity to Accra) over doing nothing as simulated through the evolution of 1000 draws from an ERGM model tracked through 1000 units of social time.}
 \label{fig:percImprovement}
\end{figure}

In Figure \ref{fig:percImprovement}, we see how the percentage improvement of the greedy strategy as compared to doing nothing changes over time for all five cities.  Error shading is $\pm$2 standard deviations as computed in a two-sample z-test with different population variances.  It is interesting to note that some cities have increasing relative improvement over time and others have decreasing relative improvement over time.   This indicates that the endogenous change mechanism, which is calibrated to each city, can reward specific changes more than others over time.

\subsection{EntrepEco: Interpretation}
  
If policy makers are interested in making the network of connections in entrepreneurial systems more like Accra's, the key finding from this study is that they should boost the connections between professionals, entrepreneur's personal networks, and startup incubators and in some cases also increasing the links between these actors and investors. There are many possible policy approaches that could be taken to achieve this. Perhaps one of the most effective approaches would simply be to disseminate these findings to startup incubators. Since startup incubators benefit directly from a vibrant entrepreneurial ecosystem, it would be in their self interest to promote entrepreneur's connections with professionals and the entrepreneur's social networks. Government could also take more direct action by using tax incentives to encourage startup incubators to focus on startups that have the help of professionals and are supported by an entrepreneur's social network. Refining these types of policy proposals will require further data collection. In future work with more cases , we will empirically calibrate combinations of metrics by using them to predict ecosystem level outcomes (e.g. \% of companies surviving past 1 year, total firms in market, market capitalization, rate of growth etc). This is an important additional step, as the structure of Accra's networks is not necessarily the cause of its success as an ecosystem.

\section{Discussion}

The framework outlined in this paper helps both to situate the existing work on network interventions but also opens up many new possibilities for intervening in networks. Through the tools of optimization, a network can potentially be changed in order to promote any goal that has a connection to the structure of a network. Our applied examples show that this approach can improve on previous approaches to disrupting networks but also that it can be used to help understand interventions that could improve networks such as to guide policy in developing markets.

While this framework takes account of more factors than previous network intervention methods, it will only perform as well as its components. For instance, if data on the network structure is incorrect or if the network recovery mechanism does not include important parameters, then the algorithm is likely to suggest suboptimal interventions. It is therefore important for future research to focus on careful data collection and careful empirical study of how networks respond to interventions. 

In addition to the variants we have discussed in the paper, there are several other things future work should consider. We currently only consider a single strategy at a time (e.g. just removing nodes or just adding edge weight), however, the approach can be extended to a mixed strategy by assigning relative costs to each action within an overall budget or having separate budgets for each type of action. This approach is likely to be most useful in cases where multiple types of action are plausible (for instance in work environments where a manager could assign employees to work together, fire an employee or change an employee's job title). Similarly, different actions could be assigned separates costs if, for instance, Noordin Top is hard to remove than one of his foot soldiers.

Future work should also continue to expand on empirically calibrating metrics to the ultimate outcomes we are interested in.  MRQAP models are one tool for this, but any model that can link network structure to overall outcomes could be used. 


\section*{Acknowledgements}
This research was funded in part by the U.S. Army Studies Program and the U.S. Army Research Office. The authors are participants in the Scientific Services Program administered by the Battelle Memorial Institute.

\section*{Author contributions statement}
All the authors contributed materially to the manuscript.  J.M. and J.Y. contributed equally by performing the experiments, writing the manuscript, and conducting analyses of the data.  D.E. produced the dataset and served as a subject matter expert on entrepreneurial networks in bourgeoning ecosystems. 
All authors reviewed the manuscript. 

\section*{Additional information}

\textbf{Competing financial interests} None.

\section*{Tables}

\begin{sidewaystable}
  \centering
  \scriptsize
    \begin{tabular}{p{2.5cm}p{3cm}p{4cm}p{4cm}p{4cm}}
    \toprule
    \textbf{Network Intervention} & \textbf{Classrooms} & \textbf{Investment community} & \textbf{Workplace} & \textbf{Criminal network} \\
    \midrule
    Strengthening and creating ties & Assigning children to work on a joint project & Facilitating a joint venture between two stakeholders & Assigning partners  & Having an informant introduce two criminals \\
    Weakening and breaking ties & Separating two children in a classroom & Removing support for a co-venture & Changing the command structure to avoid communication between certain positions in an organization & Sowing distrust between two criminals \\
    Removing nodes & Moving a child to a different classroom & Shutting down a particular business & Firing or transferring an employee & Arresting a criminal \\
    Adding nodes & Introducing a child from another classroom & Encouraging a new stakeholder to enter the community & Hiring a new employee & Having an informant infiltrate the network \\
    Changing covariate values & Giving a child a rank in the class & Giving a grant to a stakeholder & Promoting an employee & Put on most wanted list \\
    \multirow{2}[2]{*}{Aims of intervention} & \multicolumn{1}{l}{Increasing educational attainment} & \multicolumn{1}{l}{Increasing investment returns} & \multicolumn{1}{l}{Increasing employee retention} & \multicolumn{1}{l}{Decreasing number of violent crimes} \\
          & \multicolumn{1}{l}{Decreasing social segregation} & \multicolumn{1}{l}{Increasing new business formation} & \multicolumn{1}{l}{Increasing team output} & \multicolumn{1}{l}{Decreasing drug production} \\
    \bottomrule
    \end{tabular}%
  \caption{Applied Examples of network interventions}
    \label{tab:network_intervene}%
\end{sidewaystable}

\begin{table}[htbp]
  \centering

    \begin{tabular}{lrrl}
    \toprule
          & Estimate & Std. Error &  \\
    \midrule
    \textbf{Edge Covariates} &       &       &  \\
    Organization co-affiliation & 0.433 & 0.193 & * \\
    Shared education & 0.424 & 0.283 &  \\
    Friendship & 4.504 & 0.345 & *** \\
    Kin   & 4.171 & 0.670 & *** \\
    Shared religious institution & 3.654 & 0.776 & *** \\
    \textbf{Structural parameters} &       &       &  \\
    Edges & -5.744 & 0.327 & *** \\
    Isolates & -1.111 & 0.580 & . \\
    GWESP ($\alpha=0.25$) & 2.006 & 0.241 & *** \\
          &       &       &  \\
    \multicolumn{4}{l}{Signif. codes:  0 ``***" 0.001 ``**" 0.01 ``*" 0.05 ``." 0.1} \\
    \bottomrule
    \end{tabular}%
      \caption{ERGM model of Noordin terrorist network communications}
  \label{tab:ergm}%
\end{table}%

\begin{table}[ht]
\centering
\begin{tabular}{llccccc}
  \hline
Node A & Node B & Addis Ababa & Dar es Salaam & Kampala & Lusaka & Monrovia \\ 
  \hline
Govt. Rep & Social Network & 0 & 0 & 3 & 0 & 0 \\ 
Incubator & Social Network & 9 & 0 & 29 & 24 & 16 \\ 
Incubator & Professional & 8 & 0 & 18 & 26 & 10 \\ 
Investor & Social Network & 0 & 26 & 0 & 0 & 0 \\ 
Investor & Professional & 0 & 13 & 0 & 0 & 0 \\ 
Social Network & Professional & 33 & 11 & 0 & 0 & 24 \\ 
\end{tabular}
\caption{Edge weights for all entrepreneurial ecosystems in the EntrepEco dataset.}
\label{tab:allcitychanges}
\end{table}

\begin{table}[htbp]
  \centering

    \begin{tabular}{lccc}
    \toprule
          & Estimate & Pr($\geq b)$ \\
    \midrule
    Intercept & -6.6235 & -- & -- \\
    Communication & 1.3219 & 0.035 & *\\
    Education & 1.0567 & 0.065 & .\\
    Organization & 0.3727 & 0.147 & \\
    \multicolumn{4}{l}{Signif. codes:  0 ``***" 0.001 ``**" 0.01 ``*" 0.05 ``." 0.1} \\
    \bottomrule
    \end{tabular}%
      \caption{MRQAP table predicting the terrorist collaboration network}
  \label{tab:mrqapoutcome}%
\end{table}%

\end{document}